\documentclass[pra,twocolumn,pdflatex,showpacs,superscriptaddress]{revtex4}
\usepackage{graphicx}
\usepackage{dcolumn}
\usepackage{bm}
\usepackage{amssymb}
\usepackage{amsmath}
\usepackage{mathrsfs}
\usepackage{color}

\begin{document}

\title{Electromagnetic Shocks in the Quantum Vacuum}
\author{Hedvika Kadlecov\'{a}}
\affiliation{Institute of Physics of the ASCR, ELI--Beamlines project, Na Slovance 2, 18221, Prague, Czech Republic}
\author{Georg Korn}
\affiliation{Institute of Physics of the ASCR, ELI--Beamlines project, Na Slovance 2, 18221, Prague, Czech Republic}
\author{Sergei V. Bulanov}
\affiliation{Institute of Physics of the ASCR, ELI--Beamlines project, Na Slovance 2, 18221, Prague, Czech Republic}
\affiliation{National Institutes for Quantum and Radiological Science and Technology (QST),
Kansai Photon Science Institute, 8--1--7 Umemidai, Kizugawa, Kyoto 619--0215, Japan}
\affiliation{Prokhorov  General Physics Institute of the Russian Academy of Sciences, Vavilov Str.~38, Moscow 119991, Russia}
\date{\today}

\begin{abstract}
The interaction of two counter-propagating electromagnetic waves in a vacuum is analyzed within
the framework of the Heisenberg-Euler formalism in quantum electrodynamics. The nonlinear
electromagnetic wave in the quantum vacuum is characterized by wave steepening, subsequent generation of high order harmonics and electromagnetic shock wave formation with electron--positron pair
generation at the shock wave front.
\end{abstract}

\maketitle

\section{Introduction}
In contrast to classical electrodynamics where electromagnetic waves do not interact in a vacuum, in quantum electrodynamics (QED), the photon--photon scattering 
in a vacuum occurs via the generation of virtual electron--positron pairs resulting in vacuum polarization, Lamb shift, vacuum birefringence, Coulomb field modification, 
etc.~\cite{BLP-QED}. Off--shell photon–-photon scattering was indirectly observed in collisions of heavy ions accelerated by standard charged particles accelerators (see review article \cite{Baur} 
and in results of experiments obtained with the ATLAS detector at the Large Hadron Collider \cite{ATLASScattering}). 
Further study of the  process will allow extensions of the Standard Model to be tested, in which new particles can participate in loop diagrams \cite{Zempf,Inada}.

The increasing availability of high power lasers raises interest in experimental observation and motivates theoretical studies of such processes 
in laser-laser scattering \cite{Mourou, Marklund, DTomma, DiPizzaReview,  MonKod, King, Koga, KarbsteinShai}, scattering of the XFEL emitted photons \cite{Inada}, and the interaction of relatively long wavelength high intensity laser light with short wavelength X-ray photons \cite{Schlenvoigt2016, BaifeiShen2018, Heinzl2006, Shanghai100PW}. 

In the relatively low photon energy limit, for photon energy below the electron rest-mass energy, ${\cal E}_{\gamma}=\hbar \omega<m_e c^2$, the total photon--photon scattering cross section for non-polarized photons is proportional to the sixth power of the photon energy, 
\begin{equation}
\sigma_{\gamma-\gamma}=\left(\frac{973}{10125\pi}\right)\alpha^2 r_e^2 \left(\frac{\hbar \omega}{m_e c^2}\right)^6,
\label{estimate}
\end{equation}
reaching its maximum at $\hbar \omega\approx 1.5 m_e c^2$ and 
decreases proportionally to the inverse of the second power of the photon energy for $\hbar \omega>m_e c^2$ 
(see Ref.~\cite{BLP-QED}), where $\alpha=e^2/\hbar c\approx 1/137$ is the fine structure constant, $r_e=e^2/m_ec^2=2.82\times 10^{-13}\,{\rm cm}$ is the classical electron radius, $e$ and $m_e$ are electron electric charge and mass,  $c$ is speed of light in a vacuum and $\hbar$ is the reduced Planck constant.

From Eq.~(\ref{estimate}), it seems that by using the maximal frequency of the electromagnetic wave, we can reach a higher number of scattering events. This would be so if we assume the same number of photons in the colliding 
beams having the same transverse size for the beams with different frequencies. However, if we think of highest field amplitude and highest luminosity of the colliding photon beams, we must consider the smallest transverse size of the beams, i.e., they 
should be focused on a spot of one-lambda size, which is different for beams of different frequencies. In general, this approach corresponds to the Gerard Mourou's lambda-cube concept  \cite{Mourou2002}.

To find the number of photon--photon scattering events in the low frequency limit, we estimate the number of photons in the electromagnetic pulse with the amplitude $E$ in the $\lambda^3$ volume, 
 where $\lambda=2\pi c/\omega$ is the electromagnetic wave wavelength. The number of photons in a laser pulse is then equal to
\begin{equation}
N_{\gamma}=\frac{E^2  \lambda^3}{4 \pi \hbar \omega}.
\end{equation}

Using these relationships it is easy to find the number of scattering events per 4-volume $2\pi \lambda^3/\omega$. 
It is proportional to the scattering cross section given by Eq. (\ref{estimate}), to the product of the photon numbers in 
colliding photon bunches, and it is inverse proportional to the square of the wavelength. Assuming the equal photon numbers in colliding bunches and the equal photon frequencies we obtain
\begin{equation}
N_{\gamma-\gamma}=\sigma_{\gamma-\gamma}\frac{N_{\gamma}^2}{\lambda^2}
=\frac{973}{10125\pi}
\alpha^2\left(\frac{E}{E_S}\right)^4, \label{eq:phot-phot}
\end{equation}
where  $E_S=m_e^2c^3/e \hbar$ is the critical field of quantum electrodynamics.
It is also known as the Sauter-Schwinger electric field.
The corresponding to this field electromagnetic radiation intensity is $I_S=c E_S^2/4\pi\approx 10^{29}$W/cm$^2$. 
Finally and importantly, we observe that the number of scatterings does not depend on the electromagnetic wave frequency. It is determined by the
radiation intensity $I=cE^2/4\pi$ as 
\begin{equation}
N_{\gamma-\gamma}\propto \alpha^2 (I/I_S)^2,
\end{equation}
 i.e., the frequency independent dimensionless parameter characterizing photon-photon scattering is $\alpha (I/I_S)$, as in the case 
when photon--photon scattering is described within the framework of the approach based on the Heisenberg-Euler 
lagrangian used below.
We note that a similar (but not the same) analysis can be found in the papers  \cite{Schlenvoigt2016, BaifeiShen2018}. 

At the limit of extremely high amplitude of an electromagnetic field with a strength approaching the QED critical field  $E_S$, nonlinear modification of the vacuum refraction index via the 
polarization of virtual electron--positron pairs supports electromagnetic wave self--interaction.

The nonlinear properties of the QED vacuum have been extensively addressed in a number of publications.
The theoretical problem of non--linear effects of light propagation is considered in Ref. \cite{Bialynicka}, where they
 study photon splitting in an external field in the full Heisenberg--Euler theory. Another extensive studies can be found in
 Refs. \cite{Adler,Brezin, Ritus}. Other results on nontrivial vacua and on curved spacetimes can be found in  
 Refs. \cite{Latorre, DrumondHathrell, Shore}. The photon splitting in crossed electric and magnetic fields is considered,
  for example, in  \cite{PapanyanRitus}. Nonlinear wave mixing in cavities is analyzed in \cite{BrodinErikssonMarklund}.
   Nonlinear interaction between an electromagnetic pulse and a radiation background is investigated in
    \cite{MarklundBrodinStenflo}.
In the monograph \cite{DittrichGies}, the vacuum birefringence phenomena is described within 
the framework of the geometrical optics approximation by using unified formalism.  In the work \cite{Rosanov1993}, 
they incorporate weakest dispersion into Heisenberg--Euler theory, and in \cite{Lorenci} the approach used in 
Ref.  \cite{DittrichGies} is generalized allowing one to obtain the dispersion equation for the electromagnetic wave
 frequency and wavenumber. This process, in particular, results in decreasing the velocity of counter-propagating 
electromagnetic waves.  As well known, the co-propagating waves do not change their propagation velocity because the co-propagating photons do not interact, e.g., see Ref.~\cite{Zee}.

The finite amplitude wave interaction in the QED vacuum results in the high order harmonics generation \cite{Rosanov1993, DiP, FN, NF, Boehl}. High frequency harmonics generation can be a powerful tool to explore the physics of nonlinear QED vacuum. The highest harmonics can be used to probe the high energy region because they are naturally co-propagating and allow one to measure of QED effects in the coherent harmonic focus. High--order harmonics generation in vacuum is studied in detail in \cite{DiP, FN}.

Nonlinear properties of the QED vacuum in the long wavelength and low frequency limit are described 
by the Heisenberg-Euler Lagrangian \cite{HeisenbergEuler}, describing electromagnetic fields 
in dispersionless media whose refraction index 
depends on the electromagnetic field. In the media where the refraction index dependense on the field amplitude 
leads to the nonlinear response, the electromagnetic wave can evolve into a configuration with singularities \cite{LL-EDCM}. 

The appearance of singularities in the Heisenberg-Euler electrodynamics is noticed in Ref.~\cite{LutzkyToll}
 where a singular particular  solution of equations derived from the Heisenberg--Euler Lagrangian is obtained. 
 In Ref.~\cite{Boehl}, the wave steepening is demonstrated by numerical integration of nonlinear QED vacuum electrodynamics equations. 

In this paper we address the problem of nonlinear wave evolution in quantum vacuum using 
the low frequency and long wavelength approximation aiming at finding theoretical description 
of the electromagnetic shock wave formation in nonlinear QED vacuum.
We present and analyze an analytical solution of the Heisenberg--Euler electrodynamics equations 
for the finite amplitude electromagnetic wave counter-propagating to the crossed electromagnetic field. This
 configuration may correspond to the collision of the low--frequency very high intensity laser pulse 
with high frequency X-ray pulse generated by XFEL. The first, long-wavelength,  
electromagnetic wave is approximated by a constant
 crossed field. We derive the corresponding nonlinear field equations containing expressions for the relatively short 
 wavelength pulse.  The solution of the nonlinear field equations is found in a form of the simple wave or the Riemann
  wave. This solution describes high order harmonic generation, wave steepening and formation of the electromagnetic
   shock wave in a vacuum. We investigate these characteristics in more detail together with discussion of the shock
    wave front formation process.

\section{On The Heisenberg--Euler Lagrangian}

The Heisenberg--Euler Lagrangian is given by
\begin{equation}
\mathcal{L}=\mathcal{L}_{0}+\mathcal{L}', \label{eq:Lagrangian}
\end{equation}
where 
\begin{equation}
\mathcal{L}_{0}=-\frac{1}{16\pi}F_{\mu \nu}F^{\mu \nu}
\label{eq:claLagrangian}
\end{equation}
 is the  Lagrangian in classical electrodynamics, $F_{\mu \nu}$ is the electromagnetic field tensor 
($F_{\mu \nu}=\partial_{\mu} A_{\nu}-\partial_{\nu} A_{\mu}$), 
with $A_{\mu}$ being the 4-vector of the electromagnetic field
and $\mu=0,1,2,3$.
In the Heisenberg--Euler theory, the radiation corrections are described by $\mathcal{L}'$ on the right hand side of Eq.~(\ref{eq:Lagrangian}),  
which in the weak field approximation is given by \cite{HeHe},
\begin{align}
\mathcal{L}'&=\frac{\kappa}{4}\left\{\left(F_{\mu \nu}F^{\mu \nu}\right)^2 
+ \frac{7}{4} \left(F_{\mu \nu}\tilde F^{\mu \nu}\right)^2 \right. + \nonumber \\
&\frac{90}{315}\left. \left(F_{\mu \nu}F^{\mu \nu}\right) \left[  \left(F_{\mu \nu}F^{\mu \nu}\right)^2 +\frac{13}{16}\left(F_{\mu \nu}\tilde F^{\mu \nu}\right)^2  \right] \right\}
\label{eq:mathcalL}
\end{align}
where 
$\kappa=e^4/360 \pi^2 {m}^4$, $F_{\mu\nu}=\partial_{\mu}A_{\nu}-\partial_{\nu}A_{\mu}$ is and $\tilde F^{\mu \nu} = \epsilon^{\mu\nu\rho\sigma}F_{\rho\sigma}$ is a dual tensor to  a tensor of electromagnetic field, $F_{\mu \nu}$ 
with $\varepsilon^{\mu \nu \rho \sigma}$ being the Levi-Civita symbol in four dimensions.  

In the following text, we use the units $c=\hbar=1$, and the electromagnetic field is normalized on the QED critical field $E_{S}$.

To describe the singular solutions, we should keep the terms within the weak field approximation 
to the sixth order in the field amplitude, because the terms in the contributions of the fourth order cancel each other 
in calculation of dispersive properties of the QED vacuum.The remaining contribution is of the same order as from the Heisenberg--Euler Lagrangian expansion to the sixth order in the fields. We note that in Ref.~\cite{LutzkyToll}, 
the Heisenberg--Euler Lagrangian expansion to the fourth order was used. 

In the Lagrangian (\ref{eq:mathcalL}) the first two terms on the right hand side describe 
four interacting photons and the last two terms correspond to six photon interaction.

\section{Counter-propagating electromagnetic waves }

For the sake of brevity, we consider counter-propagating electromagnetic waves of the same polarization. 
They are given by the vector potential having one component, ${\bf A}=A {\bf e}_z$, with ${\bf e}_z$
 being a unit vector along the $z$ axis.
 
We assume that the electromagnetic field 4-potential written in the light cone coordinates, 
\begin{equation}
x_+=(x+t)/\sqrt{2},\quad
x_-=(x-t)/\sqrt{2},
\end{equation} 
equals 
\begin{equation}
A=W x_+ + a(x_+,x_-).
\end{equation} 
The term $W x_+$ 
describes the crossed electric and magnetic fields ($E_0=B_0=-W/\sqrt{2}$), 
whose Poynting vector is antiparallel to the $x$ axis:
\begin{equation}
{\bf P}=\frac{1}{4 \pi}{\bf E}\times {\bf B}=- \frac{W^2}{8\pi}{\bf e}_x.
\end{equation} 
Here ${\bf e}_x$ is a unit vector along the $x$-axis. 
In this case, the Lagrangian 
(\ref{eq:Lagrangian}) with $\mathcal{L}_0$ and $\mathcal{L}^{\prime}$  given by Eqs. (\ref{eq:claLagrangian}) and (\ref{eq:mathcalL}) takes the form 
\begin{equation}
\mathcal{L}=-\frac{1}{4\pi}\left[(W+w)u-\epsilon_2 (W+w)^2u^2-\epsilon_3 (W+w)^3u^3  \right] \label{eq:lightcone-Lagrangian}
\end{equation}
It depends on the functions $u=\partial_{x_-}a$ and $w=\partial_{x_+}a$.
%The electromagnetic field is normalized on the critical QED field, $E_S=m_e^2/e$. 
The dimensionless parameters
$\epsilon_2$ and $\epsilon_3$ in Eq.~(\ref{eq:lightcone-Lagrangian}) are equal to
\begin{align}
\epsilon_2&=2 e^2/45\pi =(2/45 \pi) \alpha, \label{eq:eps2}\\
\epsilon_3&=32 e^2/315\pi =(32/315 \pi) \alpha,
\end{align}
i.e., $\epsilon_2\approx 10^{-4}$ and $\epsilon_3\approx 2\times 10^{-4}$.

The field equations can be found by varying the Lagrangian. It yields
\begin{equation}
\partial_{x_-}(\partial \mathcal{L}/\partial u)+\partial_{x_+}(\partial \mathcal{L}/\partial w)=0.
\end{equation}
As a result, we obtain the system of equations
\begin{align}
\partial_{x_-} w-\partial_{x_+} u=&0, \label{eq:lightcone-1} \\
[1-4\epsilon_2 (W+w)u-9\epsilon_3u^2(W+w)^2&]\partial_{x_+} u  \nonumber \\
-[\epsilon_2  (W+w)^2+3\epsilon_3 u (W+w)^3&]\partial_{x_-} u \label{eq:lightcone-2}\\
-[\epsilon_2  u^2+3\epsilon_3 u^3(W+w)&]\partial_{x_+} w=0, \nonumber
\end{align}
where the first equation comes from the equality of mixed partial derivatives: 
$\partial_{x_-x_+}a=\partial_{x_+x_-}a$.

The equations (\ref{eq:lightcone-1}, \ref{eq:lightcone-2}) have a solution for which $u=0$ and $\partial_{x_-} w=0$, i.e., $w$ is an arbitrary
  function depending on the variable $x_+$. This is a finite amplitude electromagnetic wave propagating 
  from the right to the  left with a propagation velocity equal to speed of light in a vacuum. Its form does not change in time. 
  The electric and magnetic field components are equal to each other ($E=B=-w/\sqrt{2}$), i.e., 
  its superposition with the crossed electromagnetic field gives $E=B=-1/\sqrt{2}(W+w)$.
  
  Linearizing Eqs.~(\ref{eq:lightcone-1}, \ref{eq:lightcone-2}), it is easy to find expressions describing 
  the small amplitude wave for which we have
\begin{equation}
u(x_+,x_- )=u_0\left(x_- +\epsilon_2 W^2 x_+\right)
\label{eq:lin-wave-u}
\end{equation}
and
\begin{equation}
 \quad w(x_+,x_- )=\epsilon_2 W^2 u_0\left(x_- +\epsilon_2 W^2 x_+\right) +w_0(x_+). 
 \label{eq:lin-wave-w}
\end{equation}
In Eqs.~(\ref{eq:lin-wave-u}, \ref{eq:lin-wave-w}) the functions $u_0$ and $w_0$ are determined by the 
initial conditions.
The function $u(x_+,x_-)$ depends on the light-cone coordinates $(x_+,x_-)$ in combination
\begin{equation}
 \psi (x_+,x_-)=x_- +\epsilon_2 W^2 x_+. \label{eq:lin-wave-psi-pm}
\end{equation}
The wave phase $\psi$ can be rewritten as 
\begin{equation}
 \psi (x,t)=\frac{1}{\sqrt{2}}\left[x\left(1+\epsilon_2 W^2\right) -t\left(1-\epsilon_2 W^2\right)\right]. 
 \label{eq:lin-wave-psi-xt}
\end{equation}

The constant phase condition shows that the wave
propagates from the left to the  right with the speed
\begin{equation}
v_W=\frac{1-\epsilon_2 W^2}{1+\epsilon_2 W^2}\approx 1-2\epsilon_2 W^2+2 \epsilon_2^2 W^4.
 \label{eq:v0}
\end{equation}
It is less than unity, i.e., the wave phase (group) velocity is below the speed of light in a vacuum (see also Refs.~\cite{Marklund, Bialynicka, DittrichGies} and the literature cited therein). 

Measuring the phase difference between the phase of the electromagnetic pulse colliding with the counterpropagating
 wave and the phase of the pulse which does not interact with high intensity wave, it is equal to 
\begin{equation}
\delta \psi=4 \pi\frac{d}{\lambda}\epsilon_2 W^2,
 \label{eq:deltapsi}
\end{equation}
where $\lambda$ is the wavelength of high frequency pulse and $d$ is the interaction length, plays a central role in discussion of experimental verification of the QED vacuum birefringence \cite{King, Heinzl2006}. For 10 petawatt 
laser the radiation intensity can reach $10^{24}$W/cm$^2$, for which $W^2\approx 10^{-5}$. 
Taking the ratio equal to $d/\lambda\approx 10^4$, i.e. equal to the ratio between the optical and x-ray radiation
 wavelength, and using for $\epsilon_2$ the expression (\ref{eq:eps2}), we find that $\psi\approx 10^{-4}$.

\section{Nonlinear wave evolution }

To analyze the nonlinear wave evolution, we seek a self-similar solution to 
Eqs.~(\ref{eq:lightcone-1}, \ref{eq:lightcone-2}) of a simple wave (e.g., see \cite{Whitham, LL-HD, Kadomtsev}), 
in which  $w$ is considered as a function of $u$: $w(u)$. The simple wave (or Riemann wave) represents 
an exact solution of self--similar type of the nonlinear wave equations describing the finite amplitude wave 
propagating in a continous media. With this assumption, we obtain from Eqs.~(\ref{eq:lightcone-1}, \ref{eq:lightcone-2})
 the system of equations 
\begin{align}
&\partial_{x_+} u=J\partial_{x_-} u, \label{eq:lightconeK-1}\\
&\partial_{x_+} u=\nonumber\\
&\frac{(W+w)^2(\epsilon_2+3\epsilon_3(W+w)u)\partial_{x_-} u}
{1-(W+w)u[4\epsilon_2 +9\epsilon_3u(W+w)+3\epsilon_3 u^2J]-\epsilon_2  u^2J} ,
\label{eq:lightconeK-2}
\end{align}
where we express 
\begin{equation}
\partial_{x_+}w= J \partial_{x_+}u
\end{equation}
and
\begin{equation}
\partial_{x_-}w= J \partial_{x_-}u
\end{equation}
with the Jakobian  $J=dw/du$. Equations (\ref{eq:lightconeK-1}) and (\ref{eq:lightconeK-2}) are consistent provided that
the coefficients in front of $\partial_{x_-} u$ on the right hand sides are equal to each other. This condition yields an
equation for $J$:
\begin{align}
&[\epsilon_2  u^2+3\epsilon_3 (W+w) u^3]J^2-\left[1-4\epsilon_2(W+w)u \right. \nonumber \\
&\left.-9\epsilon_3u^2(W+w)^2\right]J+(W + w)^2[\epsilon_2 + 3\epsilon_3 u (W+w)]=0. 
\label{eq:dwduJ}
\end{align}
Using smallness of the parameters $\epsilon_2$ and $\epsilon_3$ and the relationship $w\approx \epsilon_2 W^2$,
which follows from Eq.~(\ref{eq:lin-wave-w}), we obtain an expression for the function $J(u)$ in the form of the power series: 
\begin{equation}
J(u)=\epsilon_2 W^2+4 \epsilon_2^2 W^3 u+3\epsilon_3W^3u+... 
\label{eq:dwdu2}
\end{equation}
Taking into account that $J=dw/du$ and integrating the r.h.s. of Eq. (\ref{eq:dwdu2}) with respect to the variable $u$
we obtain for the function $w(u)$ the following expression
\begin{equation}
w(u)=\epsilon_2 W^2 u+2 \epsilon_2 W^3 u^2+\frac{3}{2}\epsilon_3 W^3u^2+...
 \label{eq:dwdu-w}
\end{equation}
As a result, we find the electric and magnetic field components in the electromagnetic wave propagating 
from the left to the right
\begin{equation}
E=(w-u)/\sqrt{2}\approx -\sqrt{2}u(1-\epsilon_2 W^2) \label{eq:dwdu-E}
\end{equation}
and
\begin{equation}
B=(u+w)/\sqrt{2}\approx \sqrt{2}u(1+\epsilon_2 W^2) \label{eq:dwdu-B}
\end{equation}
respectively.

Substitution of this expression to the right hand side of Eq.~(\ref{eq:lightconeK-1}) results in 
\begin{equation}
\partial_{x_+} u-\left[\epsilon_2 W^2+(4 \epsilon_2^2 +3\epsilon_3)W^3 u\right]\partial_{x_-} u=0. 
\label{eq:simple-u}
\end{equation}
For the variables $x,t$ the equation for the function 
\begin{equation}
\bar u=-2(4\epsilon_2^2+3\epsilon_3)W^3 u
\label{eq:simple-baru}
\end{equation}
can be written as
\begin{equation}
\partial_{t} \bar u+\left(v_W+\bar u\right)\partial_{x} \bar u=0. 
\label{eq:simple-u-xt}
\end{equation}
with the velocity of linear wave, $v_W$, given by Eq. (\ref{eq:v0}).

A solution to this equation can be obtained in a standard manner (see Refs.~\cite{Kadomtsev, Whitham}). 
According to this solution,
the function $\bar u(x,t)$ transfers along the characterstic $x_0$ without distortion:
\begin{equation}
\bar u=\bar u_0(x_0)
\end{equation}
The characterstic equation for Eq.~(\ref{eq:simple-u-xt}) is 
\begin{equation}
\frac{dx}{d t } =v_W+\bar u \label{eq:charact1}
\end{equation}
with the solution 
\begin{equation}
x=x_0+(v_W+\bar u_0(x_0))t. 
\label{eq:charact2}
\end{equation}
Combining these relationships, we obtain the solution to Eq.~(\ref{eq:simple-u-xt}) in the implicit form,
where the function $u(x,t)$ should be found from equation
\begin{equation}
\bar u=\bar u_0\left(x-(v_W+\bar u)t\right). 
\label{eq:implicite}
\end{equation}
In particular, this expression describes high order harmonics generation and wave steepening in a vacuum. 

Various mechanisms for generating high order harmonics in the QED vacuum are analyzed in 
Refs.~\cite{Rosanov1993, DiP, FN, Boehl}. In particular, the parametric wave interaction process was considered in \cite{Rosanov1993} and the ``relativistic oscillating mirror" concept 
(for details of this concept, see \cite{ROM, ROMR, RFM2}) was applied in \cite{Boehl}. Here, we formulate perhaps one of the simplest mechanisms. To obtain the scaling of high order harmonics generation within the framework of this mechanism, we choose the initial electromagnetic wave as
\begin{equation}
\bar u_0=\bar a_1\cos(k(x-v_W t)), 
\label{eq:HoH1}
\end{equation}
where $\bar a_1$ and $k=\omega/c$ are the wave amplitude and wave number ($\omega$ is the wave frequency), respectively. Using the weakness of nonlinearity ($\bar a_1 \ll 1$), we find from Eqs.~(\ref{eq:implicite}, \ref{eq:HoH1}) that
\begin{equation}
\bar u(x,t)=\bar a_1\cos(k(x-v_W t))-\frac{\bar a_1^2}{2} k v_W t\sin(2k(x-v_W t))... 
\label{eq:HoH2}
\end{equation}
Taking into account normalization of the wave amplitude given by Eq.~(\ref{eq:simple-baru}),
we find that the ratio of the second harmonic amplitude to the amplitude of the wave with fundamental frequency scales as $(2 \epsilon_2^2 W^3+3\epsilon_3W^3/2)\bar a_1 k v_W t$.
It is proportional to the duration of the electromagnetic wave interaction. Assuming $kv_W t=2\pi d/\lambda$ 
as in the case corresponding to Eq. (\ref{eq:deltapsi}), and the intensity of the x-ray pulse of $10^{21}$W/cm$^2$
we obtain that the ratio is approximately 
equal to $10^{-11}$.

From expression (\ref{eq:implicite}), it follows that the electromagnetic field gradient increases with time, i.e., wave steepening occurs. Differentiating $u(x,t)$ with respect to the coordinate $x$, we find
\begin{equation}
\partial_x u=\frac{\partial_{x_0}u_0(x_0)}{1-2(4\epsilon_2^2+3\epsilon_3)W^3  \partial_{x_0}u_0(x_0) t}, \label{eq:grad1}
\end{equation}
where the dependence of the Lagrange coordinate $x_0$ on time and the Euler coordinate $x$ is given by 
Eq.~(\ref{eq:charact2}). As shown, the gradient $\partial_x u$ becomes infinite at time 
\begin{equation}
t_{br}=\frac{1}{2(4\epsilon_2^2+3\epsilon_3)W^3 | \partial_{x_0}u_0(x_0) |} 
\label{eq:tbr1}
\end{equation}
and at the coordinate $x_0$ where the derivative $\partial_{x_0}u_0(x_0) $  has its maximum. 
This singularity is called the `` gradient catastrophe'' or ``the wave breaking''. 

\begin{figure}[h]
\centering
\includegraphics[width=0.5\textwidth]{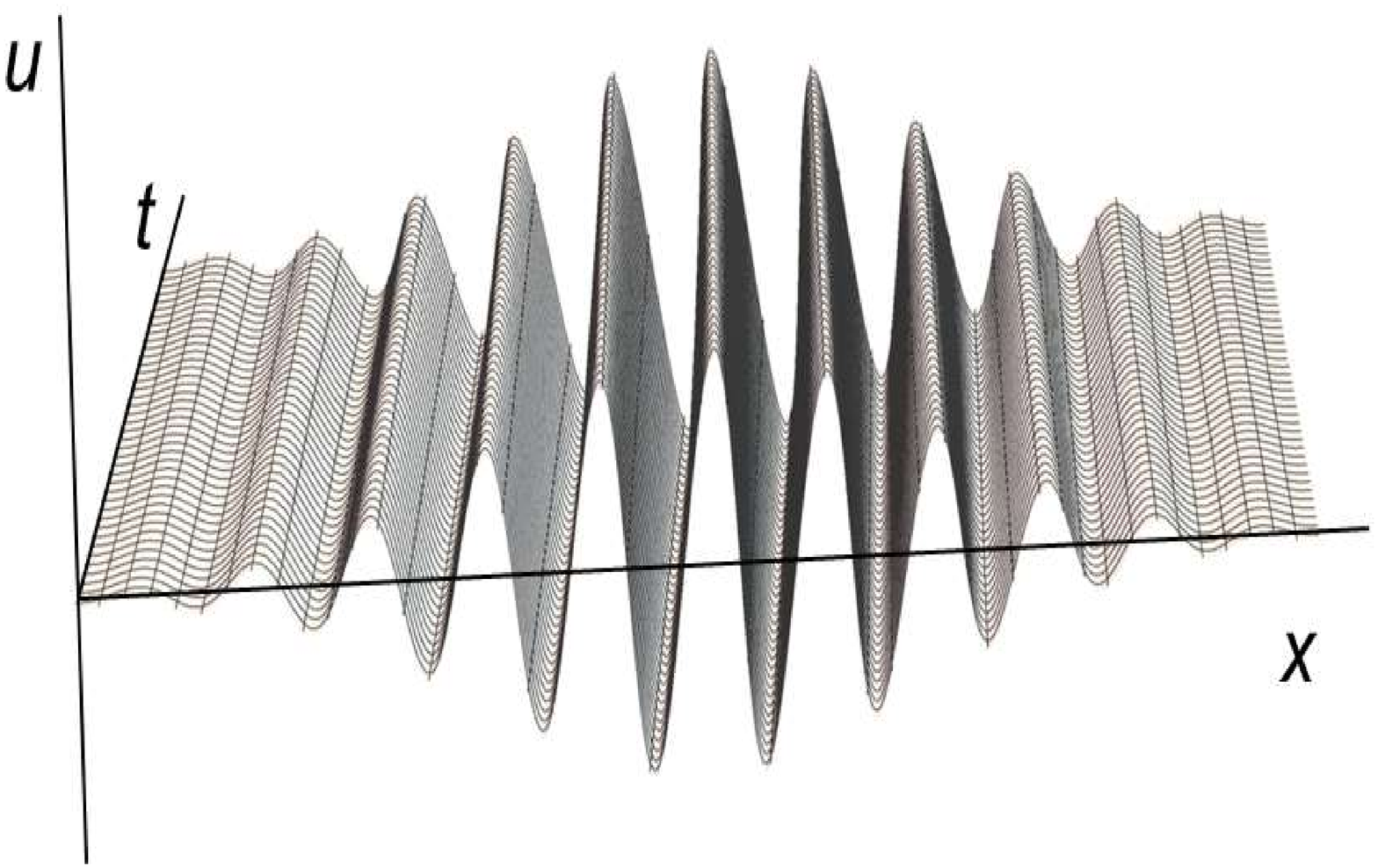}
\caption{\label{Fig1} 
The function $u(x,t)$ given by Eq.~(\ref{eq:u0x0}) for $u_{0}(x_{0})$ given by the expression (\ref{eq:u0x0}). }
\end{figure}

\begin{figure}[h]
\centering
\includegraphics[width=0.43\textwidth]{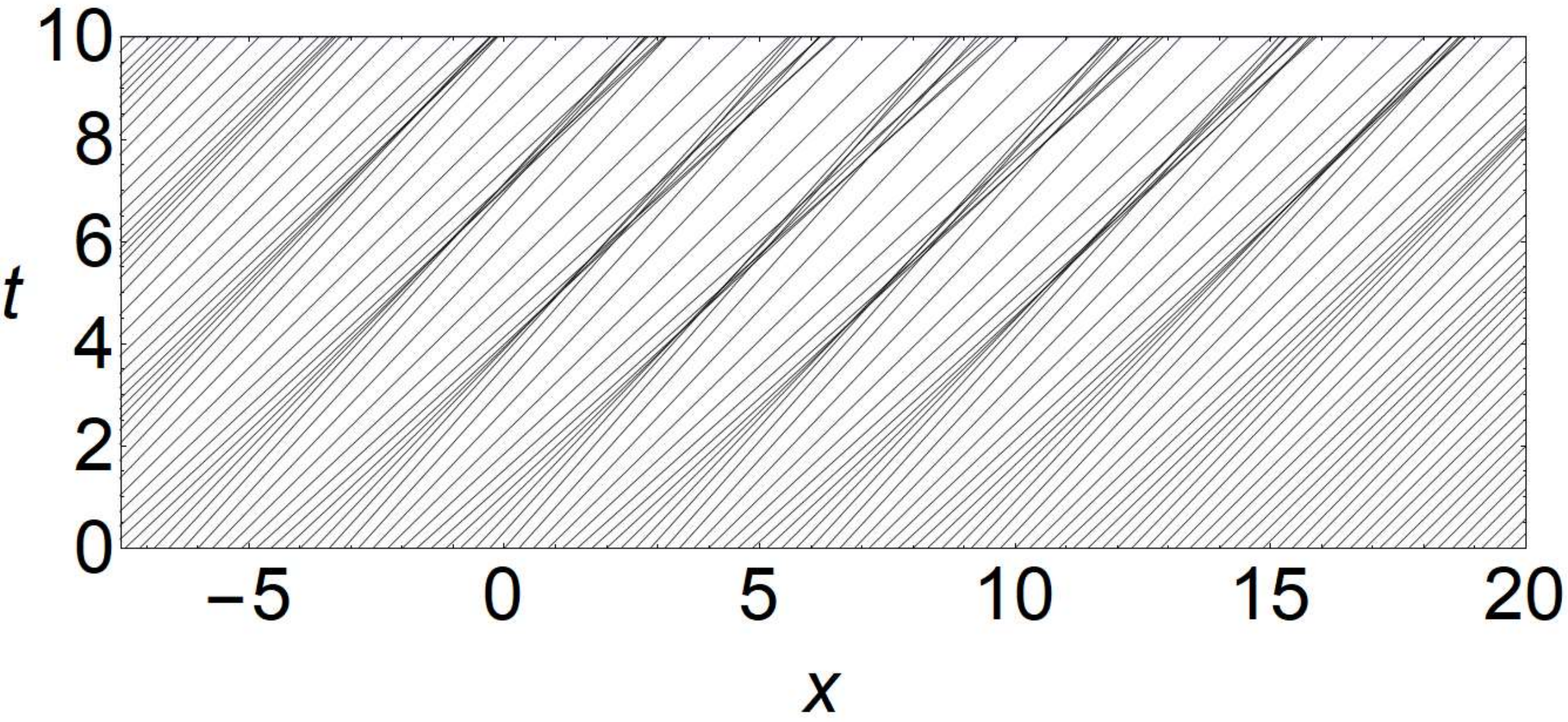}
\caption{\label{Fig2} 
Characteristics of Eq.~(\ref{eq:implicite}) plotted in the $x, t$ plane for the same parameters as in Fig.~(\ref{Fig1}).}
\end{figure}

The formation of singularity during the evolution of a finite amplitude electromagnetic wave in the quantum vacuum
is illustrated in Figures \ref{Fig1} and \ref{Fig2}. The electromagnetic pulse at $t=0$ takes the form
\begin{equation}
u_0(x_0)=a_0 \exp (-x_0^2/2 L^2)\cos (k x_0),\label{eq:u0x0}
\end{equation}
where $L=4\pi$ and $k=2$. The parameter $4 \epsilon_{2}^2 W^3+3\epsilon_3W^3$ is assumed to be equal to 0.125 and $a_0=1$. As clearly shown 
in Fig.~\ref{Fig1}, wave steepening evolves with time. Wave breaking occurs due to the characteristic intersection as shown in Fig.~\ref{Fig2}.

We note that the singularity formed at the electromagnetic wave breaking corresponds to 
the rarefaction shock wave formation (the wave steepens and breaks in the backwards direction, as shown in Fig.~1), because the wave crest propagates with a speed lower than the propagation speed of the part of the
pulse with lower amplitude.

When a wave approaches the wave breaking point, wave steepening is equivalent to harmonics generation with higher numbers. Because of this fact the long-wavelength approximation used above becomes 
inapplicable and the Heisenberg-Euler Lagrangian cannot be used to describe the wave evolution in the vicinity of the gradient catastrophe, i.e., at the shock wave front.
According to the shock wave paradigm \cite{LL-HD}, the basic properties of the shock wave (the relationships between the shock wave velocity and the average parameters of the medium before and after the shock wave front) can be found within the framework in “the long-wavelength” approximation if we consider the shock wave front region as a discontinuity. 

\section{Electromagnetic shock wave in vacuum }

The long-wavelength approximation breaks when the frequencies of 
the interacting waves, $\omega_{\gamma}$ and $\Omega$ become high enough, i.e., when their product becomes of the order or higher than 
\begin{equation}
\omega_{\gamma} \Omega>m_e^2 c^4/\hbar^2.
\label{Threshold}
\end{equation}

At this photon energy level the photon--photon 
interaction can result in the creation of real electron--positron pairs during the Breit-Wheeler process \cite{BW}, 
in the saturation of
wave steepening, and in electromagnetic shock wave formation. Here, $\omega_{\gamma} $ and 
$\Omega$ are the frequencies of high energy photons and low frequency counter-propagating electromagnetic waves, respectively.

The electromagnetic shock wave separates two regions, I and II, where the function $u(x,t)$ takes 
the values $u_I$ and $u_{II}$, respectively. 
The shock wave front (it is an interface between regions I and II) moves with a velocity equal to $v_{sw}$, 
in other words, the shock wave front is localized at the position $x_{sw}=v_{sw}t$.
Integrating Eq.~(\ref{eq:simple-u-xt}) over an infinitely small interval $(-\delta+x_{sw}, x_{sw}+\delta)$,
where $\delta \to 0$, we obtain
\begin{equation}
\{-v_{sw}u+v_W u -(4\epsilon_2^2+3\epsilon_3)W^3u^2\}_{x=x_{sw}}=0.
\label{eq:discont}
\end{equation}
Here, $\{f\}_x=f(x+\delta)-f(x-\delta)$ at $\delta\to 0$ denotes the discontinuity 
of the function $f(x)$ at the point $x$. From Eq.~(\ref{eq:discont}),
it follows that 
\begin{equation}
v_{sw}=v_W -(4\epsilon_2^2+3\epsilon_3)W^3 (u_I+u_{II}).
\label{eq:velshock}
\end{equation}
Near the threshold $\omega \Omega\geq m_e^2 c^4/\hbar^2$, the electron--positron creation cross section 
equals \cite{BLP-QED,PinG-G}
\begin{equation}
\sigma_{e-p}=\pi r_e^2 \sqrt{\frac{\hbar^2 \omega \Omega}{m_e^2 c^4}-1}.
\label{eq:sigma-e-p}
\end{equation}

The width of the shock wave front can be estimated to
be of the order of the length $l_{sw} = 1/n_{\gamma} \sigma_{e-p}$, over which
the photon with energy of the order of $m_ec^2/\hbar$ creates the electron--positron pair. 
The photon density $n_{\gamma}$
is related to the electromagnetic pulse energy ${\cal E}_{em}$ as
$n_{\gamma}\approx{\cal E}_{em}/\hbar \omega \lambda^3 N_{em}$, where $N_{em}=l_{em}/\lambda$ 
is the electromagnetic pulse length divided by the wavelength. It yields
$l_{sw} \approx \hbar \omega \lambda^3 N_{em}/\pi r_e^2 {\cal E}_{em}$. Since it is assumed that the
photon energy is at the threshold of the electron-positron
pair creation, the shock wave front width should be of the order
of the Compton wavelength, $\lambdabar_C=\hbar/m_ec$. 
This condition imposes a constraint from below on the electromagnetic pulse energy of ${\cal E}_{em}\geq m_ec^2(\lambda/r_e)^2$. For a 1$\mu$m
wavelength laser with $N_{em} = 10$, it requires ${\cal E}_{em}\geq100$kJ.
If $\lambda=10^{-8}$cm$^2$, which corresponds to an X-ray pulse of 10 KeV,
we have ${\cal E}_{em}\geq 10^{-4}$J.

The electron-positron pairs created at the electromagnetic shock wave front being accelerated by the 
electromagnetic wave emit gamma-ray photons which lead to the
electron--positron avalanche via the multi-photon Breit-Wheeler mechanism \cite{AIN-VIR} as discussed in 
Refs. \cite{BellKirk, FAM} (see also review article
\cite{DiPizzaReview} and the literature cited therein). This process requires the dimensionless parameters
$\chi_e\approx(E/E_S)\gamma_e$ and $\chi_{\gamma}\approx(E/E_S)(\hbar \omega_{\gamma}/m_ec^2)$ 
to be greater than one. At $\chi_{\gamma}>1$, the QED vacuum becomes a dispersive and dissipative 
media \cite{NBN1969, Erber1966}. The effects of the electromagnetic wave dispersion  originate formally 
from the high order derivatives in the 
corrections to the Heisenberg-Euler Lagrangian found in Refs. \cite{Mamayev1981, gusyninI, gusyninII}. 
Discussions of higher-order derivatives when describing the QED vacuum beyond the 
Heisenberg-Euler Lagrangian model can be found in Refs. \cite{Mamayev1981} and \cite{gusyninI,gusyninII}.
As it has been noted in Refs. \cite{SS-2000, RNN1998} the implementation of the high derivatives 
into the description of nonlinear wave interaction in QED vacuum can result in the soliton formation. 
In \cite{RNN1998} it is noted that the counterplay of nonlinearity 
and dispersion in nonlinear QED vacuum can lead to the dark soliton formation, which can be interpreted as the 
shock wave of the electromagnetic pulse envelope. Details on the dark soliton properties can be found in Refs. \cite{DS1,DS2,DS3} and the literature cited therein.

In our case, the dispersion can result in modulations of the electromagnetic field in the vicinity of the shock wave front.  
Whether the dispersive properties come into play well below the pair-production threshold (\ref{Threshold}) depends not
 only on the wave amplitudes but also on the frequency of the interacting waves. For example, in the case of $10$ KeV
  X-ray radiation, the dispersive effects prevail at an intensity above $10^{26}$ $W/cm^2$. In any case, we do not
   expect this would change the main result of our paper because dissipation/dispersion determines the shock front
    structure. 

With additional terms with derivatives in the Heisenberg--Euler Lagrangian (\ref{eq:Lagrangian}), the  theory
predicts also the existence of bright spatial solitons \cite{SS-2000}.  With the presence of dispersion, the resulting 
   wave break is prevented and it results in the process when the first soliton is formed. 
   This process continues until the initial pulse is completely splitted into the chain of the solitons 
   (see for more details Refs. \cite{GP, SPN}). Such scenario needs huge peak
    intensity $10^{33} W/cm^2$ which  can be decreased for experimental observation by making the size of the soliton
     large compared to the carier  wavelength. For $\lambda=10\,{\rm nm}$ the peak intensity is $10^{25}\,W/cm^2$ 
      \cite{SS-2000}.

\section{Conclusion }

In conclusion,
we presented and analyzed an analytical solution of the Heisenberg--Euler electrodynamics equations describing the
 finite amplitude electromagnetic wave counter-propagating to the crossed electromagnetic field.  The solution belongs 
 to the family of self-similar solutions corresponding to the Rieman wave.
It describes the wave steepening and formation of the electromagnetic shock wave in the vacuum and the 
high order harmonic generation. 

The singularity formed at the electromagnetic wave breaking has rarefaction shock wave character  
(the wave steepens and breaks in the backwards direction, as illustrated in Fig.~1), because the wave 
crest propagates with a speed lower than the propagation speed of the part of the pulse with lower amplitude.

In general, photon--photon scattering in a vacuum
is governed by the dimensionless parameter $\alpha (I_{em}/I_S)$,
as it concerns shock-like configuration formation, high order harmonics generation and the electron-positron and gamma ray flash at the electromagnetic
shock wave front. Observation of these phenomena in a high power laser or x-ray interaction with matter
implies high precision measurements as in experiments \cite{Baur, ATLASScattering} or achieving an electromagnetic field amplitude approaching the critical QED field $E_S$. 
One of the ways of reaching these regimes is to increase the laser power.
For example, observation of one scattered photon per day with a 1 Hz laser requires 
an intensity of the order of $8\times 10^{27}$W/cm$^2$, i.e., several hundred kJ 
laser energy.
Another way of approaching the critical QED field limit is
associated with the relativistic flying mirror  concept
\cite{Koga} (for relativistic flying mirror theory and experiments see Refs.~\cite{RFM1, RFM2, RFM3, RFM4}),
 where light intensity can be increased during the nonlinear laser-plasma interaction.

\begin{acknowledgments}
We thank Drs.~T. Heinzl, R. Sauerbrey, N. N. Rosanov, J. Nejdl, T. Esirkepov, and J. Koga for productive discussions.
The work is supported by the project: High Field Initiative (CZ$.02.1.01/0.0/0.0/15\_003/0000449$) under the European Regional Development Fund.
\end{acknowledgments}

%\bibliographystyle{unsrt} 
%\bibliography{/home/thiago/bibtex/articles,/home/thiago/bibtex/books}

%\end{document}

%\bibliography{apssampELI1}
% Produces the bibliography via BibTeX.

\end{document}